\documentclass[aps,prd,reprint,amsmath,amssymb,superscriptaddress,nofootinbib]{revtex4-2}

\usepackage[T1]{fontenc}
\usepackage{graphicx}
\usepackage{amssymb}
\usepackage{amsthm,amsmath}
\usepackage{subfig}
\usepackage{placeins}
\usepackage[dvipsnames]{xcolor}

\usepackage[colorlinks=true,
            linkcolor=Red,
            citecolor=Red,
            urlcolor=Red]{hyperref}
\begin{document}

\title{Probing a heavy neutral lepton in a radiative neutrino model at a muon collider}

\author{Isaac Bamwidhi}
\email{isaac.bamwidhi@uaeu.ac.ae}
\email{isaac.bamwidhi@cern.ch}

\author{Mohamed Belfkir}
\email{mohamed.belfkir@cern.ch}

\author{Mohamed Amin Loualidi}
\email{ma.loualidi@uaeu.ac.ae}

\author{Salah Nasri}
\email{snasri@uaeu.ac.ae}
\email{salah.nasri@cern.ch}

\affiliation{Department of Physics, United Arab Emirates University, Al Ain, United Arab Emirates}
\affiliation{College of General Education, University of Doha for Science and Technology, Doha, Qatar}

\begin{abstract}
We investigate direct and Higgs-mediated production of a Dirac heavy neutral lepton in a radiative neutrino-mass model at a future $3~\mathrm{TeV}$ muon collider. We consider heavy neutral lepton masses in the window $m_Z<m_N<m_h$, where the decays $N\to\ell W$ and $N\to\nu Z$ can proceed with on-shell weak bosons. We consider both direct production, $\mu^+\mu^-\to N\nu$, and Higgs-mediated production through $\mu^+\mu^-\to\nu_\mu\bar\nu_\mu h$, $h\to N\nu$. Both channels lead to the common $\ell^\pm jj+p_{\mathrm T}^{\mathrm{miss}}$ final state. We perform a detector-level analysis with an integrated luminosity of $1~\mathrm{ab}^{-1}$, assuming unpolarized muon beams, and derive median expected $95\%$ confidence-level upper limits on the visible signal cross section ranging from approximately $0.027$ to $0.113~\mathrm{fb}$, corresponding to limits on the mixing strength parameter $|U_{\mu N}|^2$ ranging from $6.7\times10^{-6}$ to $8.3\times10^{-5}$ for six benchmark points. The Higgs-mediated production channel contributes $\sim15$--$72\%$ of the signal after the common event selection, but improves the expected limits by $\sim0.18$--$1.19\%$ within the common classifier and binning scheme. This small improvement reflects the classifier-score distribution of the Higgs-mediated events. The limits on $|U_{\mu N}|^2$ account for the benchmark-specific flavor mixings and assume an independent $10\%$ background uncertainty in each classifier-score bin.

\end{abstract}

\maketitle

\section{Introduction}
\label{introduction}
Neutrino oscillations establish that at least two light neutrinos are massive, requiring physics beyond the Standard Model (SM). Radiative neutrino-mass models explain the smallness of these masses by generating them through loop processes. The loop suppression can accommodate new particles near the electroweak scale, making them potentially accessible to collider searches. Among the new states that can appear in such models, heavy neutral leptons (HNLs) are particularly interesting because their production and decay can be probed experimentally. For comprehensive reviews of heavy neutral leptons (HNLs) and radiative neutrino-mass models, see Refs.~\cite{Abdullahi:2022jlv,Cai:2017jrq}. Many collider searches for HNLs focus on production via mixing with active neutrinos, followed by leptonic or semileptonic decay signatures at hadron and lepton colliders~\cite{Atre:2009rg,Deppisch:2015qwa,Banerjee:2015gca}. For HNL masses below the Higgs-boson mass, Higgs decays provide an additional production channel, while Higgs precision measurements offer complementary constraints~\cite{Das:2017zjc,Das:2017rsu,Bernal:2023coo}. More recent studies have treated the Higgs--HNL coupling and active--heavy neutrino mixing as independent parameters \cite{Bernal:2024kqx}. In neutrino mass models, however, the underlying Yukawa structure can relate these quantities, motivating a model-consistent treatment of HNL production and decay. 

A concrete realization is the \texttt{T4-3-i} topology, a one-loop completion of the dimension-five Weinberg operator involving scalar and fermionic mediators~\cite{Bonnet:2012kz,Belfkir:2026slp}. In Ref.~\cite{Bamwidhi:2026T43iC1}, we studied the corresponding \texttt{T4-3-i-C1} model in detail. Its particle content includes an inert scalar doublet, a real scalar triplet, a Majorana fermion triplet, and a Dirac fermion singlet $N$. An exact $Z_2$ symmetry forbids the tree-level type-III contribution and stabilizes the lightest $Z_2$-odd particle, which can provide a scalar dark matter candidate. After electroweak symmetry breaking, the Yukawa interaction between $N$, the SM lepton doublets, and the Higgs doublet induces mixing between $N$ and the active neutrinos. It also gives rise to an $hN\nu$ vertex. When $m_N<m_h$, the Higgs boson can therefore decay through $h\to N\nu$, where the charge-conjugate channel is also understood, providing a direct collider probe of this interaction.

In this paper, we investigate the sensitivity of the Muon Collider (MuC), which is proposed as a future high-energy lepton collider with a possible first stage operating at $\sqrt{s}=3~\mathrm{TeV}$ and delivering an integrated luminosity of $1~\mathrm{ab}^{-1}$~\cite{InternationalMuonCollider:2024jyv,MuonCollider:2022xlm,Delahaye:2019omf,AlAli:2021let,Black:2022cth,Narain:2022qud,Aime:2022flm}, to the gauge singlet Dirac HNL of the \texttt{T4-3-i-C1} model. To this end, we consider six benchmark points in the window $m_Z<m_N<m_h$, with HNL masses in the range $m_N\sim103\text{--}119~\mathrm{GeV}$, using their corresponding couplings and branching fractions. The signal receives contributions from direct associated production, $\mu^{+}\mu^{-}\to N\nu$, and Higgs-mediated $WW$-fusion production, $\mu^{+}\mu^{-}\to\nu_{\mu}{\bar\nu_{\mu}}h$ with $h\to N\nu$. In this mass range, the decays $N\to\ell^{\pm}W^{\mp}$ and $N\to\nu Z$ can proceed through on-shell weak bosons. Assuming prompt HNL decay, both production channels lead to an $\ell^{\pm}jj+p_{\mathrm T}^{\mathrm{miss}}$ signature. We perform a detector-level analysis based on the expected performance of a multipurpose muon collider detector~\cite{deFavereau:2013fsa,MuonCollider:2022ded}. Motivated by the work presented in Refs.~\cite{Li:2023tbx,Mekala:2023diu,Kwok:2023dck,Thor:2023nzu,Knolle:2024eys,Chen:2025nzn},  we train an \textsc{XGBoost} classifier using reconstructed kinematic and angular observables associated with the leading charged lepton, the hadronically decaying  $W$ boson, the HNL candidate, the missing transverse momentum ($p_{\mathrm{T}}^{\mathrm{miss}}$), and the overall event topology. We train the classifier on the combined Direct and $WW$-fusion signals and the relevant SM backgrounds, and evaluate its performance on an independent weighted test sample. We then incorporate the binned classifier response into a profile likelihood $\mathrm{CL_s}$ analysis to derive the expected $95\%$ confidence-level limits on the visible signal cross section. We interpret these limits in terms of the active-heavy mixing parameter $|U_{\mu N}|^2$. The distinguishing feature of this study is its combined treatment of direct and Higgs-mediated HNL production below the Higgs mass threshold within a common model-consistent benchmark framework. Because their contributions vary across benchmark points, we retain both production modes with their corresponding couplings, normalizations, and kinematic distributions throughout the analysis. Although the Higgs-mediated channel contributes 15\%--72\% of the selected signal, it improves the expected limits by only 0.18\%--1.19\% within the common classifier and binning scheme. This is because its events populate classifier-score regions with comparatively lower sensitivity.

We summarize the theoretical framework in Sec.~\ref{sec:theoretical_framework}. Section~\ref{sec:signal_simulation} describes the collider analysis, while Sec.~\ref{sec:results} presents the results and Sec.~\ref{sec:conclusion} the conclusion.
\section{Theoretical framework}
\label{sec:theoretical_framework}
As in the full \texttt{T4-3-i-C1} model described in Ref.~\cite{Bamwidhi:2026T43iC1}, we retain a single generation of Dirac singlets $N=N_L+N_R$, with a rank-two one-loop light-neutrino mass matrix, leaving one massless active neutrino. The interaction relevant to HNL production and decay is
 \begin{equation}
  \mathcal{L}_{Y_N} =\sum_{\alpha=e,\mu,\tau}(Y_N)_\alpha \overline{N_R} \widetilde H^{\dagger}L_\alpha + \mathrm{h.c.}, \quad \widetilde H\equiv i\sigma_2 H^* .
  \label{YN}
\end{equation}
Here $L_\alpha$ and $H$ are the SM lepton and Higgs doublets.  After electroweak symmetry breaking, this interaction generates the Dirac mixing masses  $(m_D)_\alpha=v(Y_N)_\alpha/\sqrt{2}$ and the Higgs vertex
\begin{equation}
    \mathcal{L}_{hN\nu}= \sum_\alpha g_\alpha h\overline{N}P_L\nu_\alpha+\mathrm{h.c.},
    \qquad g_\alpha=\frac{(Y_N)_\alpha}{\sqrt{2}}.
    \label{eq:hNnu_interaction}
\end{equation}
In the small-mixing limit $|(m_D)_\alpha|\ll M_N$, where $M_N$ is the singlet Dirac mass, the physical  HNL mass satisfies $m_N\simeq M_N$, and the active-HNL mixing parameters are given, to leading order, by
\begin{equation}
\begin{aligned}
    |U_{\alpha N}|^2&\simeq\frac{|(m_D)_\alpha|^2}{m_N^2},
    \qquad |U_N|^2\equiv\sum_\alpha|U_{\alpha N}|^2,\\
    g_{\mathrm{eff}}^2&\equiv\sum_\alpha|g_\alpha|^2
    \simeq\frac{m_N^2}{v^2}|U_N|^2.
\end{aligned}
\label{eq:effective_coupling}
\end{equation}
For $m_N<m_h$, the notation $h\to N\nu$ includes both charge-conjugate final states $N\bar\nu_\alpha$ and $\bar N\nu_\alpha$. Tree-level partial-wave unitarity constraints on the total HNL Yukawa coupling have been derived in Ref.~\cite{Urquia-Calderon:2024rzc}, and are already satisfied by our benchmark couplings. Summing over flavors and these final states, and neglecting light-neutrino masses, we obtain
\begin{equation}
    \Gamma(h\to N\nu)=\frac{m_h}{8\pi}\,g_{\mathrm{eff}}^2
    \left(1-\frac{m_N^2}{m_h^2}\right)^2.
    \label{eq:hNnu_width}
\end{equation}
Assuming no other appreciable non-SM Higgs partial widths, the corresponding branching fraction is
\begin{equation}
    \mathrm{Br}(h\to N\nu)=
    \frac{\Gamma(h\to N\nu)}{\Gamma_h^{\mathrm{SM}}+\Gamma(h\to N\nu)}.
    \label{eq:BR_h_to_Nnu}
\end{equation}
We use the Standard Model Higgs-boson width $\Gamma_h^{\mathrm{SM}}=4.070~\mathrm{MeV}$~\cite{LHCHiggsCrossSectionWorkingGroup:2016ypw}. The same active-heavy mixing also induces couplings of $N$  to the $W$ and $Z$ bosons, with the relevant interactions at leading order in $U_{\alpha N}$ given by  
\begin{equation}
\begin{aligned}
    \mathcal{L}_{W,Z}\supset{}&
    \frac{g}{\sqrt{2}}W^-_\mu\sum_\alpha U_{\alpha N}
    \overline{\ell_\alpha}\gamma^\mu P_L N\\
    &+\frac{g}{2c_W}Z_\mu\sum_\alpha U_{\alpha N}
    \overline{\nu_\alpha}\gamma^\mu P_L N+\mathrm{h.c.},
\end{aligned}
\label{eq:HNL_weak_currents}
\end{equation}
where $g$ is the $SU(2)_L$ gauge coupling and $c_W=\cos\theta_W$. We restrict the benchmarks to $m_Z<m_N<m_h$. In this window, the two-body decays $N\to\ell_\alpha^-W^+$ and $N\to\nu Z$ proceed through on-shell weak bosons, while $N\to\nu h$ is kinematically forbidden. Neglecting charged-lepton masses and defining $r_V=m_V^2/m_N^2$, the corresponding partial widths are
\begin{equation}
\begin{aligned}
    \Gamma(N\to\ell_\alpha^-W^+)&=
    \frac{g^2m_N^3}{64\pi m_W^2}|U_{\alpha N}|^2
    (1-r_W)^2(1+2r_W),\\
    \Gamma(N\to\nu Z)&=
    \frac{g^2m_N^3}{128\pi c_W^2m_Z^2}|U_N|^2
    (1-r_Z)^2(1+2r_Z).
\end{aligned}
\label{eq:HNL_partial_widths}
\end{equation}
Here $\Gamma(N\to\nu Z)$ is summed over light-neutrino mass eigenstates.  The signal branching fractions for $N\to\ell W$ are calculated using the total HNL width $\Gamma_N$, including $N\to\nu Z$ and any other open decay channels. Since 
the inclusive Higgs and neutral-current decay widths depend on $|U_N|^2$, whereas each charged-current partial width depends on $|U_{\alpha N}|^2$, interpreting the collider limits in terms of  $|U_{\mu N}|^2$ requires a specified flavor-mixing pattern applied consistently to production and decay. 

To verify the validity of the  prompt-decay approximation for our benchmark points,  we estimate an upper bound on the mean HNL decay length in the laboratory frame, starting from the proper mean decay length determined by the total HNL width, $c\tau_N=\hbar c/\Gamma_N$. Since the mean decay length in the laboratory frame is enhanced by the HNL boost, we obtain a conservative upper bound by considering  the maximum kinematically allowed boost at the collider,
\begin{equation}
    (\beta\gamma)_{\mathrm{max}} =\frac{s-m_N^2}{2\sqrt{s}m_N},
    \qquad L_{\mathrm{max}}^{\mathrm{mean}}
    =(\beta\gamma)_{\mathrm{max}}c\tau_N.
\end{equation}
For the benchmark points, $L_{\mathrm{max}}^{\mathrm{mean}}$ ranges from  $\sim2.8\times10^{-10}$ to $6.0\times10^{-9}~\mathrm{m}$, well below typical displaced vertex resolution scales. The prompt-decay approximation is thus justified, and Table~\ref{tab:hnl_decay_lengths} gives the detailed values.

\begin{table}[htbp]
    \centering
   \begin{ruledtabular}
        \begin{tabular}{cccc}
            $m_N$ [GeV] &
            $\Gamma_N$ [GeV] &
            $c\tau_N$ [m] &
            $L_{\max}^{\mathrm{mean}}$ [m] \\
            \hline
            103.089 & $4.817115\times10^{-7}$ & $4.10\times10^{-10}$ & $5.95\times10^{-9}$ \\
            103.355 & $3.656084\times10^{-6}$ & $5.40\times10^{-11}$ & $7.82\times10^{-10}$ \\
            109.727 & $9.778775\times10^{-6}$ & $2.02\times10^{-11}$ & $2.75\times10^{-10}$ \\
            112.174 & $2.194180\times10^{-6}$ & $8.99\times10^{-11}$ & $1.20\times10^{-9}$ \\
            114.322 & $4.248084\times10^{-6}$ & $4.65\times10^{-11}$ & $6.09\times10^{-10}$ \\
            119.088 & $1.023969\times10^{-6}$ & $1.93\times10^{-10}$ & $2.42\times10^{-9}$ \\
        \end{tabular}
    \end{ruledtabular}
    \caption{Total HNL widths, proper decay lengths, and maximum mean laboratory-frame decay lengths for the six benchmark points at $\sqrt{s}=3~\mathrm{TeV}$.}
    \label{tab:hnl_decay_lengths}
\end{table}
\section{COLLIDER ANALYSIS}
\label{sec:signal_simulation}
We implemented the \texttt{T4-3-i-C1} model in \textsc{FeynRules} 2.3.49~\cite{Alloul:2013bka} and exported it to the Universal FeynRules Output (UFO) format~\cite{Degrande:2011ua}, including the active HNL mixing and interactions required for the production and prompt decay of the HNL.
\subsection{Event generation}
\label{sec:event_generation}
We consider unpolarized muon--antimuon collisions at a center-of-mass energy of $\sqrt{s}=3~\mathrm{TeV}$ and generate the signal and background samples using \textsc{MadGraph5\_aMC@NLO}~\cite{Alwall:2014hca}. We incorporate initial-state radiation (ISR), at the hard event generation stage using the built-in \textsc{MadGraph} lepton-ISR treatment, while the equivalent photon approximation is used for the relevant photon-initiated processes. The simulated samples are normalized to an integrated luminosity of $1~\mathrm{ab}^{-1}$. The signal events were generated with six HNL masses $m_N=103.089, 103.355, 109.727, 112.174, 114.322$, and $119.088~\mathrm{GeV}$. The model scan contains two additional points in the interval $95<m_N<100~\mathrm{GeV}$, at $m_N=95.188$ and $99.919~\mathrm{GeV}$. Although both satisfy the prompt decay requirement, their direct production cross-sections of $1.78\times10^{-7}$ and $2.50\times10^{-8}~\mathrm{pb}$, respectively, correspond to fewer than $0.2$ events before detector selections. We therefore do not include them as additional benchmarks. Table~\ref{tab:branching_ratios} summarizes the total Higgs width, the partial width for $h\to N\nu$, and its branching fraction for the six HNL mass hypotheses. 
\begin{table}[htbp]
\centering
\resizebox{\columnwidth}{!}{%
\begin{tabular}{cccc}
\hline\hline
$m_N$ [GeV] &
$\Gamma_{h}$ [GeV] &
$\Gamma(h\to N\nu)$ [GeV] &
$\mathrm{BR}(h\to N\nu)$ \\
\hline
$103.089$ & $4.0702893\times10^{-3}$ & $2.8928\times10^{-7}$ & $7.1071\times10^{-5}$ \\
$103.355$ & $4.0721097\times10^{-3}$ & $2.1097\times10^{-6}$ & $5.1809\times10^{-4}$ \\
$109.727$ & $4.0720589\times10^{-3}$ & $2.0589\times10^{-6}$ & $5.0562\times10^{-4}$ \\
$112.174$ & $4.0702971\times10^{-3}$ & $2.9715\times10^{-7}$ & $7.3004\times10^{-5}$ \\
$114.322$ & $4.0703717\times10^{-3}$ & $3.7165\times10^{-7}$ & $9.1307\times10^{-5}$ \\
$119.088$ & $4.0700241\times10^{-3}$ & $2.4089\times10^{-8}$ & $5.9186\times10^{-6}$ \\
\hline\hline
\end{tabular}%
}
\caption{Total width of the SM-like Higgs boson, partial width
for $h\to N\nu$, and corresponding branching fraction for the
six HNL benchmark points.}
\label{tab:branching_ratios}
\end{table}
{The benchmark flavor mixing parameters and HNL branching fractions used in the signal simulation are summarized in Table~\ref{tab:flavor_benchmarks}.
\begin{table*}[htbp]
    \centering
    \begin{tabular}{c c c c c c c}
        \hline\hline
        $m_N$ [GeV]
        & $|U_{eN}|^2$
        & $|U_{\mu N}|^2$
        & $|U_{\tau N}|^2$
        & $\mathrm{Br}(N\to eW)$
        & $\mathrm{Br}(N\to\mu W)$
        & $\mathrm{Br}(N\to\tau W)$ \\
        \hline
        103.089
        & $1.9804\times10^{-6}$
        & $7.7240\times10^{-7}$
        & $4.8461\times10^{-7}$
        & 0.522113 & 0.203627 & 0.127319 \\
        103.355
        & $1.0615\times10^{-5}$
        & $5.7729\times10^{-6}$
        & $7.6698\times10^{-6}$
        & 0.375599 & 0.204267 & 0.270472 \\
        109.727
        & $1.9110\times10^{-6}$
        & $1.6723\times10^{-5}$
        & $2.0963\times10^{-5}$
        & 0.038651 & 0.338240 & 0.423053 \\
        112.174
        & $3.9523\times10^{-6}$
        & $2.3413\times10^{-6}$
        & $1.3005\times10^{-6}$
        & 0.409163 & 0.242378 & 0.134380 \\
        114.322
        & $9.2922\times10^{-7}$
        & $6.0595\times10^{-6}$
        & $5.9693\times10^{-6}$
        & 0.055628 & 0.362756 & 0.356740 \\
        119.088
        & $3.0773\times10^{-7}$
        & $8.0206\times10^{-7}$
        & $1.3173\times10^{-6}$
        & 0.095902 & 0.249953 & 0.409940 \\
        \hline\hline
    \end{tabular}
    \caption{HNL mixing parameters and branching fractions for the six benchmark points.}
    \label{tab:flavor_benchmarks}
\end{table*}
For each benchmark point, we generate separate signal samples for the direct and the $WW$-fusion production modes.
For the direct production mode, we have
\begin{equation}
    \resizebox{\columnwidth}{!}{%
    $
    \mu^{+}\mu^{-}
    \to N\nu
    \to \ell^{\pm}W^{\mp}\nu
    \to \ell^{\pm}q\bar q'\nu
    \to \ell^{\pm}jj+p_{\mathrm{T}}^{\mathrm{miss}},
    \qquad \ell=e,\mu.$%
    }
    \label{eq:direct_signal_chain}
\end{equation}\\
and for the $WW$-fusion production mode we have 
\begin{equation}
\begin{aligned}
\mu^{+}\mu^{-}
&\to \nu_{\mu}\bar{\nu}_{\mu}h
 \to \nu_{\mu}\bar{\nu}_{\mu}N\nu
 \to \nu_{\mu}\bar{\nu}_{\mu}\ell^{\pm}W^{\mp}\nu \\
&\to \nu_{\mu}\bar{\nu}_{\mu}\ell^{\pm}q\bar{q}'\nu
 \to \ell^{\pm}jj+p_{\mathrm{T}}^{\mathrm{miss}}.
\end{aligned}
\label{eq:wwfusion_signal_chain}
\end{equation}

Both production mechanisms lead to the same reconstructed $\ell^{\pm}jj+p_{\mathrm{T}}^{\mathrm{miss}}$ final state. The two modes differ in their invisible-particle content: direct production contains one associated neutrino, whereas the Higgs-mediated channel contains additional neutrinos from $WW$ fusion and $h\to N\nu$. We generate the complete $\mu^+\mu^-\to\nu_\mu\bar{\nu}_\mu h$ matrix element. The sample is dominated by $WW$ fusion, while the Higgsstrahlung contribution, $\mu^+\mu^-\to Zh$ with
$Z\to\nu_\mu\bar{\nu}_\mu$ only accounts for $\sim 0.024\%$ of the generated events. We retain these events. For each HNL mass hypothesis, $550000$ events are generated separately for both production modes. Each simulated sample is normalized using the event weight
\begin{equation}
    w_i=\frac{\sigma_i\,\mathcal{L}}{N_i^{\mathrm{gen}}},
    \label{eq:event_weight}
\end{equation}
where $\sigma_i$ denotes the \textsc{MadGraph} cross section for the generated process, $\mathcal{L}$ is the integrated luminosity, and $N_i^{\mathrm{gen}}$ is the corresponding number of generated events in sample $i$. Since the event generation includes the decay chains, the corresponding branching fractions are already accounted for and are not applied separately.

For both production modes, the signal is characterized by one charged lepton, at least two jets associated with the hadronic decay of the $W$ boson, and  $p_{\mathrm{T}}^{miss}$ carried by neutrinos. The relevant Standard Model
backgrounds are therefore processes that produce or mimic the
$\ell^{\pm}jj+p_{\mathrm{T}}^{\mathrm{miss}}$ final state. The simulated background categories are
\begin{align}
    \mathrm{B01}:&\quad
    \mu^{+}\mu^{-}
    \to \mu^{+}\nu_{\mu}W^{-},
    \ \mu^{-}\bar{\nu}_{\mu}W^{+}
    \to \mu^{\pm}\nu_{\mu}^{(-)}jj ,
    \nonumber\\[2pt]
    \mathrm{B02}:&\quad
    \mu^{+}\mu^{-}
    \to \mu^{+}\mu^{-}Z
    \to \mu^{+}\mu^{-}jj ,
    \nonumber\\[2pt]
    \mathrm{B03}:&\quad
    \gamma\gamma
    \to W^{+}W^{-}
    \to \mu^{\pm}\nu_{\mu}^{(-)}jj ,
    \nonumber\\[2pt]
    \mathrm{B04}:&\quad
    \mu^{+}\mu^{-}
    \to \nu_{\mu}\bar{\nu}_{\mu}
    \mu^{\pm}\nu_{\mu}^{(-)}W^{\mp}
    \to \nu_{\mu}\bar{\nu}_{\mu}
    \mu^{\pm}\nu_{\mu}^{(-)}jj ,
    \nonumber\\[2pt]
    \mathrm{B05}:&\quad
    \mu^{+}\mu^{-}
    \to W^{+}W^{-}
    \to e^{+}\nu_{e}jj ,
    \nonumber\\[2pt]
    \mathrm{B06}:&\quad
    \gamma\gamma
    \to W^{+}W^{-}
    \to e^{\pm}\nu_{e}^{(-)}jj ,
    \nonumber\\[2pt]
    \mathrm{B07}:&\quad
    \mu^{+}\gamma
    \to \bar{\nu}_{\mu}W^{+}Z
    \to \bar{\nu}_{\mu}jj\,e^{+}e^{-},
    \nonumber\\
    &\quad
    \gamma\mu^{-}
    \to \nu_{\mu}W^{-}Z
    \to \nu_{\mu}jj\,e^{+}e^{-},
    \nonumber\\[2pt]
    \mathrm{B08}:&\quad
    \mu^{+}\mu^{-}
    \to \nu_{\mu}\bar{\nu}_{\mu}
    e^{\pm}\nu_{e}^{(-)}W^{\mp}
    \to \nu_{\mu}\bar{\nu}_{\mu}
    e^{\pm}\nu_{e}^{(-)}jj .
    \label{eq:background_processes}
\end{align}
Where $\nu_{\ell}^{(-)}$ denotes the neutrino or antineutrino required by the associated lepton charge. The background set is restricted to the dominant electroweak processes containing a prompt electron or muon and a hadronically decaying vector boson. Contributions from $t\bar t$, leptonic $\tau$ decays, and other electroweak processes are not included. We generate $1.2\times10^{6}$ events for the dominant $W\mu\nu$ background process B01 and $2\times10^{5}$ events for each of the remaining seven background processes. For the $\mathrm{B07}$, we only generate $\mu^+\gamma\to\bar{\nu}_\mu W^+Z$. The charge conjugate is included by multiplying its normalization by a factor of two. 

After hard-scattering event generation, we pass both the signal and background events to \textsc{Pythia~8}~\cite{Sjostrand:2014zea} for parton showering and hadronization, and model the detector response with \textsc{Delphes}~\cite{deFavereau:2013fsa} using a dedicated muon collider detector card. We use the reconstructed photons, electrons, muons, jets, tracks, and the $p_\mathrm{T}^{miss}$ obtained from the \textsc{Delphes} output to define the analysis objects and event selection requirements described in Sec.~\ref{sec:objects}. Beam-induced backgrounds are not explicitly overlaid in the detector simulation. Furthermore, we do not include separate detector response or object misidentification uncertainties.
\subsection{Objects and event selection}
\label{sec:objects}
The baseline analysis objects are charged leptons, jets, and $p_{\mathrm T}^{\mathrm{miss}}$, reconstructed from the simulated detector response. Tracks enter only through the charged-lepton isolation variable. We describe the definitions and selection criteria applied to these objects below.

\paragraph{Tracks}
Tracks are taken from the \textsc{Delphes} track collection. They are required to satisfy $p_{\mathrm{T}} > 0.5~\mathrm{GeV}$ and $|\eta| < 2.5$ as determined by the detector card parametrization tracking efficiency. We impose no additional track-level selection in the analysis.

\paragraph{Photons}
Photon candidates are taken from the \textsc{Delphes} photon collection. No additional photon-level selection is applied in the analysis. 

\paragraph{Electrons}
Electron candidates are taken from the \textsc{Delphes} electron collection.  They are required to satisfy $p_{\mathrm{T}} \geq 5~\mathrm{GeV}$ and $|\eta| < 2.5$. The highest-$p_{\mathrm{T}}$ candidate among the selected central electrons is taken to be the leading lepton. No explicit isolation requirement is imposed at this stage. 

\paragraph{Muons}
Central muon candidates are taken from the \textsc{Delphes} muon collection and are required to satisfy $p_{\mathrm{T}} \geq 5~\mathrm{GeV}$ and $|\eta| < 2.5$. Forward muons are required to satisfy $p_{\mathrm{T}} \geq 5~\mathrm{GeV}$ and $2.5 < |\eta| < 8.0$ and are only considered if there is no central electron or muon. No isolation requirement is imposed.

The relative lepton-isolation variable is calculated from energy-flow objects with $p_{\mathrm{T}} > 0.5~\mathrm{GeV}$ within a cone of $\Delta R = 0.1$ around the lepton. We retain the isolation variable for the subsequent multivariate analysis.

\paragraph{Jets}
Jets are reconstructed with the inclusive Valencia (VLC) algorithm and are required to satisfy $p_{\mathrm{T}} > 10~\mathrm{GeV}$ and $|\eta| \leq 2.5$. The hadronically decaying $W$-boson candidate is formed from the jet pair whose invariant mass is closest to the nominal $W$-boson mass. Candidates reconstructed with $R=0.7$ are preferred, while those with $R=0.2$ are retained if no resolved $R=0.7$ candidate is available. The selected $W$-boson candidate is required to satisfy $\Delta R\left(\ell,W\right) < 1.5$

\paragraph{Missing transverse momentum}
The missing transverse momentum, $p_{\mathrm{T}}^{\mathrm{miss}}$, is obtained from the \textsc{Delphes} energy-flow objects. We keep its magnitude and azimuthal angle for the multivariate analysis. No extra requirement is imposed. 

The cumulative effect of the selection requirements is summarized in Table~\ref{tab:cutflow}.
\begin{table}[htbp]
    \centering
        \resizebox{\columnwidth}{!}{%
        \begin{tabular}{lrrrr}
        \hline\hline
        Selection
        & Direct & $WW$ fusion & $W\mu\nu$ & Other backgrounds \\
        \hline
        All events
        & 445.2 & 53.95 & $2.709\times10^{6}$ & $2.081\times10^{6}$ \\
        
        Lepton baseline
        & 382.2 & 50.64 & $2.587\times10^{6}$ & $1.937\times10^{6}$ \\
        
        Resolved $\ell jj$ candidate
        & 30.82 & 28.90 & $1.515\times10^{6}$ & $2.364\times10^{5}$ \\
        
        $\Delta R(\ell,W)<1.5$
        & 28.87 & 17.65 & $1.558\times10^{5}$ & $1.700\times10^{4}$ \\
        
        \hline\hline
        \end{tabular}%
        }
        \caption{
        Expected event yields at $\sqrt{s}=3$ TeV and $\mathcal{L}=1~\mathrm{ab}^{-1}$ after baseline selections for the representative $m_N=109.727$ GeV benchmark and the principal backgrounds. The 'other backgrounds' column combines background samples B02--B08.
        }
        \label{tab:cutflow}
\end{table}

Figure~\ref{fig:kinematics1} shows the normalized distributions of the reconstructed hadronic-(W) candidate mass, $m_{jj}$, and the HNL-candidate mass, $m_{\ell jj}$, after the baseline selection for the $m_N=109.727~\mathrm{GeV}$ benchmark.
\begin{figure*}[htbp]
    \centering
    \includegraphics[width=0.90\linewidth]{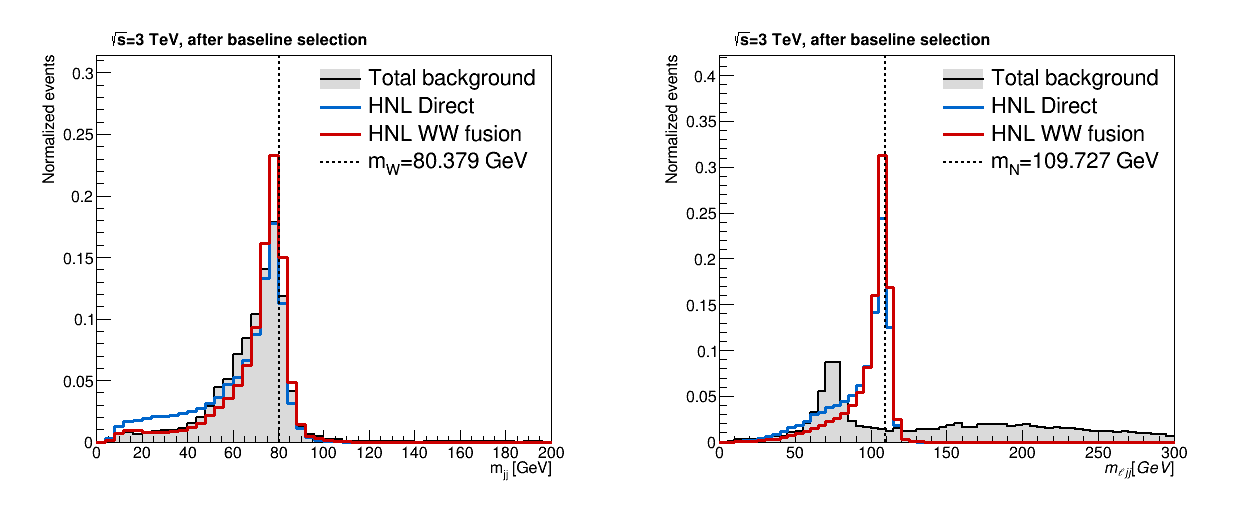}
    \caption{Normalized distributions of the signals and total backgrounds}
    \label{fig:kinematics1}
\end{figure*}

\subsection{Classifier and event weights}
\label{sec:ml_event_categorization}
We require that, in addition to the baseline selections described in Sec.~\ref{sec:objects}, the selected events have a leading isolation variable below $0.3$. The direct and $WW$-fusion samples for all six mass hypotheses are then combined and assigned the signal label, while the eight SM processes B01--B08 are combined and assigned the background label. The two signal contributions remain separately normalized before their
expected yields are added. The direct and Higgs-mediated samples retain their separate event weights and classifier-score distributions. We add their expected yields, bin by bin, only when constructing the combined signal model used in the likelihood. We then split the resulting dataset randomly into two statistically independent training and test subsets. The training dataset contains $75\%$ of the contribution, and the test dataset contains $25\%$. This is done through \textsc{scikit-learn}~\cite{Pedregosa:2011ork}. During training, each event is assigned a combined weight $w_{\mathrm{train},i}=w_{\mathrm{event},i}w_{\mathrm{class}}(y_i)$, where $w_{\mathrm{event},i}=w_i/\mathcal{L}$ is the cross-section weight corresponding to the sample normalization defined in Eq.~\eqref{eq:event_weight}, and $y_i\in\{0,1\}$ is the background or signal label. For each class $c$, the class weight is calculated as 
\begin{equation}
    w_{\mathrm{class}}(c)=\frac{\displaystyle\sum_i w_{\mathrm{event},i}}
    {\displaystyle N_{\mathrm{classes}}\sum_{i:y_i=c}w_{\mathrm{event},i}},
    \qquad N_{\mathrm{classes}}=2.
\end{equation}
This equalizes the total effective weights of the signal and background classes in the selected dataset. The physical event weights $w_i$ determine the expected sample yields, while the additional class weights $w_{\mathrm{class}}$ are used only during classifier training to balance the total effective signal and background weights. No normalization or scaling is applied to the input variables.

\subsection{Tree-based classifier}
\label{subsec:classifier}
We use the \textsc{XGBoost} framework~\cite{Chen:2016btl,Shanahan:2022ifi} to construct a binary classifier that separates the combined direct and $WW$-fusion signals from the SM backgrounds. We train this classifier using reconstructed detector-level observables that describe the leading lepton, the hadronic $W$-boson candidate, the reconstructed HNL candidate, $p_{\mathrm{T}}^{\mathrm{miss}}$, and the overall event topology. We select the variables to exploit the characteristic $N\to\ell W\to\ell jj$ decay structure without using generator-level truth information. The complete set of variables is summarized in Table~\ref{tab:xgb_inputs}.
\begin{table}[htbp]
    \centering
    \resizebox{\columnwidth}{!}{%
    \begin{tabular}{ll}
    \hline\hline
        \textbf{Category} & \textbf{Input variables} \\
        \hline
        Lepton category
        & $C_e$, $C_\mu$, $C_{\mu_{\mathrm{fwd}}}$ \\
        Leading lepton
        & $p_{\mathrm{T}}^\ell$, $|\eta_\ell|$,
          $I_{\mathrm{rel}}^\ell$ \\
        Jet candidate
        & $R_{\mathrm{cand}}$, $N_{\mathrm{jet}}^{R_{\mathrm{cand}}}$ \\
        Hadronic $W$ candidate
        & $p_{\mathrm{T}}^W$, $|\eta_W|$, $m_{jj}$ \\
        HNL candidate
        & $\Delta R(\ell,W)$, $m_{\ell jj}$,
          $p_{\mathrm{T}}^{\ell jj}$ \\
        Missing transverse momentum
        & $p_{\mathrm{T}}^{\mathrm{miss}}$,
          $|\Delta\phi(\ell,\mathbf{p}_{\mathrm{T}}^{\mathrm{miss}})|$,
          $|\Delta\phi(W,\mathbf{p}_{\mathrm{T}}^{\mathrm{miss}})|$ \\
        Event composition
        & $N_{\ell}^{\mathrm{noiso+fwd}}$ \\
        \hline\hline
    \end{tabular}%
    }
    \caption{Reconstructed detector-level observables used as inputs
    to the \textsc{XGBoost} classifier.}
    \label{tab:xgb_inputs}
\end{table}
Here, $C_e$, $C_\mu$, and $C_{\mu_{\mathrm{fwd}}}$ are binary category variables for a leading central electron, central muon, and forward muon, respectively. For each selected event, the variable corresponding to the leading lepton category is set to $ 1$, and the other two are set to $0$. The candidate radius $R_{\mathrm{cand}}$ specifies whether the preferred resolved $W$ candidate is reconstructed with $R=0.7$ or, when no such candidate is available, constructed with $R=0.2$. $N_{\mathrm{jet}}^{R_{\mathrm{cand}}}$ denotes the VLC-jet multiplicity at the selected radius and $N_{\ell}^{\mathrm{noiso+fwd}}$ is the sum of the central electron and muon multiplicities before isolation and the forward-muon multiplicity.

We optimize the \textsc{XGBoost} hyperparameters using a randomized search over the parameter space~\cite{Pedregosa:2011ork}. The combined event and class weights are also applied during the optimization. We then use the best-performing parameter set to train the final classifier on the training subset. The reserved test subset is used only for the final performance evaluation. The optimized hyperparameters are summarized in Table~\ref{tab:xgb_hyperparameters}.
\begin{table}[htbp]
    \centering
    \resizebox{\columnwidth}{!}{%
    \begin{tabular}{llr}
    \hline
        Parameter & Definition & Optimized value \\
        \hline
        \texttt{colsample\_bytree} & Fraction of input variables sampled per tree & 0.876 \\
        \texttt{gamma} & Minimum loss reduction required for a split & 0.456 \\
        \texttt{learning\_rate} & Boosting step size & 0.0714 \\
        \texttt{max\_delta\_step} & Maximum change in a leaf weight & 5.757 \\
        \texttt{max\_depth} & Maximum tree depth & 7 \\
        \texttt{min\_child\_weight} & Minimum child-node instance weight & 9 \\
        \texttt{n\_estimators} & Number of boosting trees & 981 \\
        \texttt{reg\_alpha} & $L_1$ regularization coefficient & 0.382 \\
        \texttt{reg\_lambda} & $L_2$ regularization coefficient & 8.790 \\
        \texttt{scale\_pos\_weight} & Internal positive-class weighting factor & 1.729 \\
        \texttt{subsample} & Fraction of training events sampled per tree & 0.245 \\
        \hline
    \end{tabular}%
    }
    \caption{Optimized \textsc{XGBoost} hyperparameters used in the final
    classifier.}
    \label{tab:xgb_hyperparameters}
\end{table}

The trained classifier assigns each event a score between zero and one. Larger values correspond to events with more signal. Fig.~\ref{fig:xgb_performance} compares the weighted classifier-score distributions for the training and independent test samples and the receiver operating characteristic curve (\textsc{ROC}) evaluated on the test sample. The weighted test curve gives an area of $0.970$ under the curve. 
\begin{figure}[htbp]
    \centering
    \includegraphics[width=0.46\textwidth]{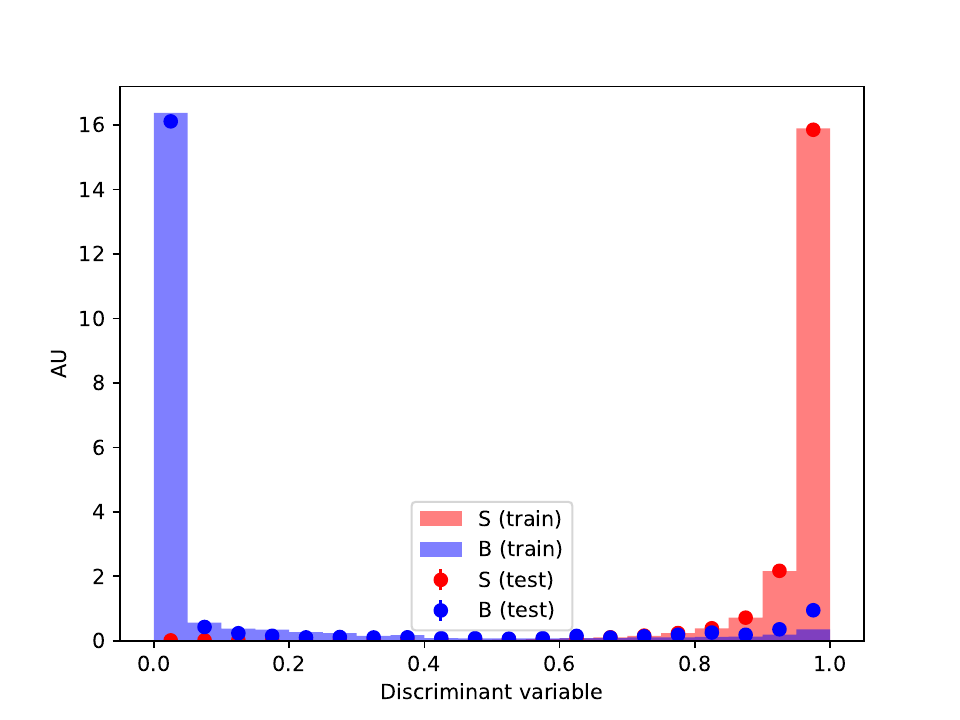}

    \includegraphics[width=0.40\textwidth]{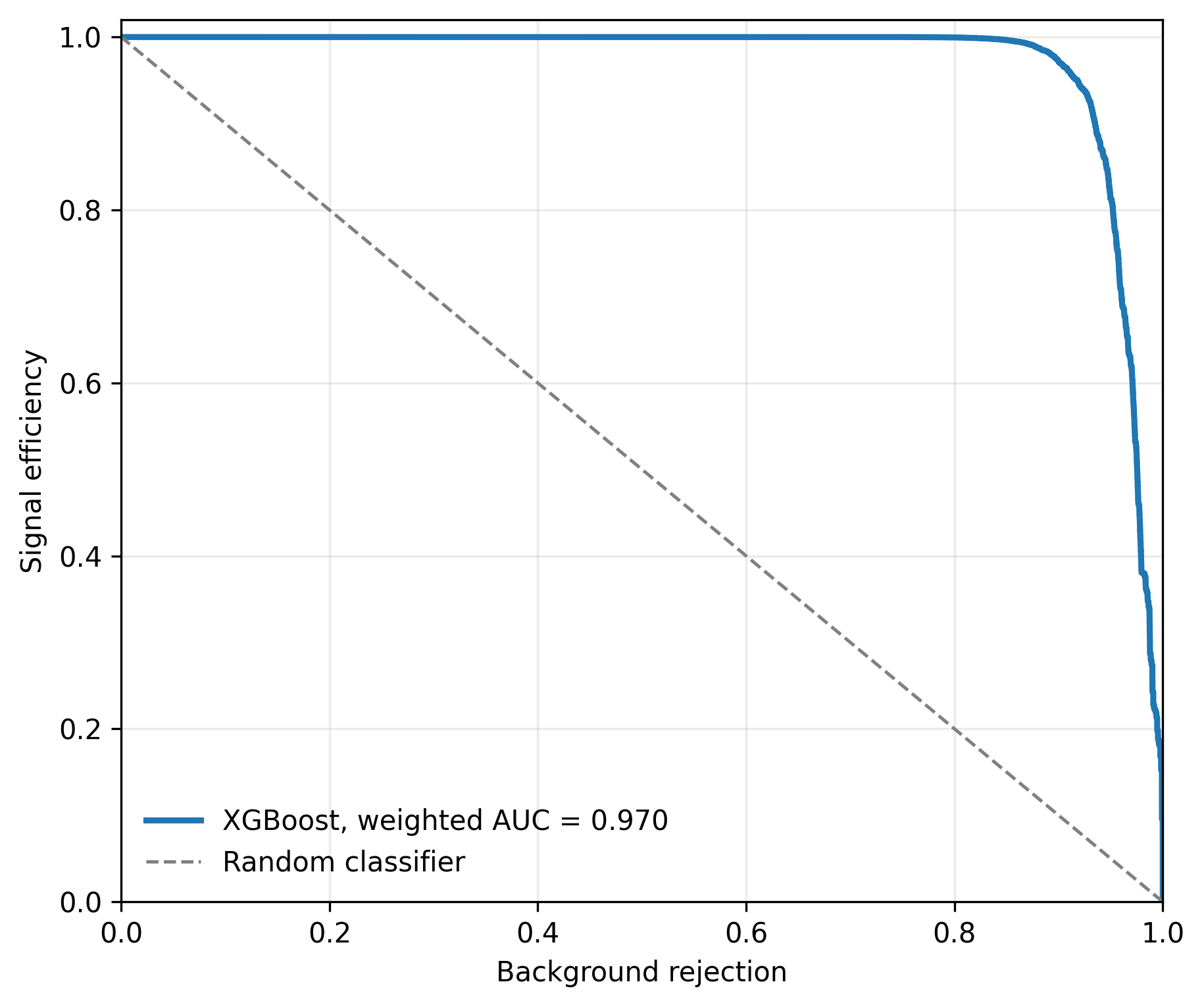}
    \caption{\textsc{XGBoost} classifier performance. Top: weighted
    classifier-score distributions for the combined signal and
    background samples in the training and independent test
    subsets. The test-sample points include statistical
    uncertainties. Bottom: weighted ROC curve evaluated on the test
    sample, with a weighted area under the curve of $0.970$.}
    \label{fig:xgb_performance}
\end{figure}

We evaluate the input variables using \textsc{SHAP} (SHapley Additive exPlanations) values~\cite{lundberg2017unified}. The variables are ordered in Fig.~\ref{fig:shap} according to their mean absolute \textsc{SHAP} values. The reconstructed HNL mass provides the largest overall contribution to the classifier response.
\begin{figure}[htbp]
    \centering
    \includegraphics[width=0.7\linewidth]{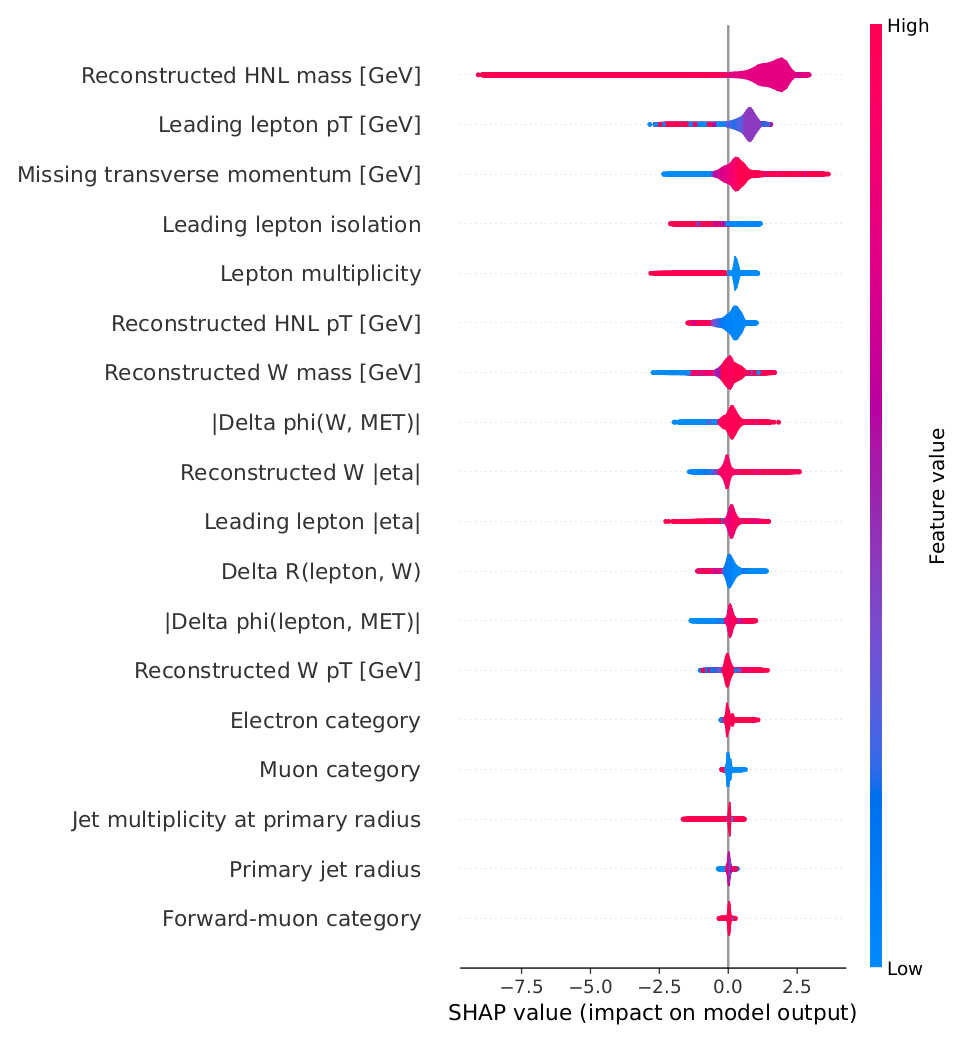}
    \caption{\textsc{SHAP} summary for the trained \textsc{XGBoost} classifier. The
    input variables are ordered by decreasing mean absolute \textsc{SHAP}
    value. Each point represents an event; the horizontal position
    shows the effect of the variable on the classifier output, while
    the color indicates the corresponding variable value. Positive
    \textsc{SHAP} values shift the prediction toward the signal class and
    negative values toward the background class.}
    \label{fig:shap}
\end{figure}

\section{Results and statistical analysis}
\label{sec:results}
For the statistical analysis, the classifier score distribution is divided into four bins. We fix the boundary between the lowest score interval and the signal-enriched region at $0.5$. The remaining two boundaries are determined separately for each HNL mass hypothesis by maximizing the \textsc{Asimov} significance~\cite{Cowan:2010js},
\begin{equation}
    Z_{\mathrm{A}}=\sqrt{2\left[(S+B)\ln\left(1+\frac{S}{B}\right)-S\right]},
\end{equation}
where $S$ and $B$ are the expected signal and background yields, respectively. The two boundaries are sequentially optimized. First, we maximize $Z_{\mathrm A}$ for the highest score interval and then apply the same criterion to the interval below it. We combine the four resulting intervals only in the subsequent likelihood fit. During the boundary optimization, candidate intervals with an expected background yield below one event or with zero expected signal yield are discarded. Once determined, we fix the bin boundaries and apply them, unchanged, to the independent $25\%$ test subset. The test event weights are rescaled by $1/0.25$ to recover the expected yields at the full integrated luminosity. This independent test subset is used to construct the binned likelihood. We obtain the likelihood using the \textsc{pyhf}~\cite{Feickert:2022lzh} framework. The signal yield in all bins is scaled by a common signal strength parameter $\mu$. $\mu=1$ corresponds to the nominal signal prediction and $\mu=0$ corresponds to the background only hypothesis. We assign an uncorrelated $10\%$ uncertainty to the expected background yield in each bin. We estimate the finite simulation uncertainties in each process and score bin as $\delta N_{\mathrm{MC}}=\sqrt{\sum_i w_i^2}$. However, we do not include them in the nominal likelihood. We instead assess the sparsely sampled $W\mu\nu$ contribution in the highest score bin by assigning a conservative upper bound to its yield as described below. We calculate expected upper limits at the $95\%$ confidence level using the asymptotic $\mathrm{CL_s}$ method~\cite{Read:2002hq}. We then evaluate the median expected limit and the corresponding $\pm 1\sigma$ and $\pm 2\sigma$ bands under the background-only hypothesis. To assess this dependence, we repeat the fit with uncorrelated background uncertainties of $5\%$ and $20\%$ in each classifier score bin, while retaining $10\%$ as the nominal assumption. The median expected upper limits on the signal strength for the three uncertainties are 
\begin{equation*}
    \begin{array}{c|ccc}
        m_N~[\mathrm{GeV}]
        & \mu_{95}(5\%)
        & \mu_{95}(10\%)
        & \mu_{95}(20\%) \\
        \hline
        103.089 & 8.5378   & 8.6271   & 8.7769   \\
        103.355 & 1.5266   & 1.5436   & 1.5710   \\
        109.727 & 1.9954   & 2.0681   & 2.1554   \\
        112.174 & 3.2653   & 3.2908   & 3.3563   \\
        114.322 & 6.3491   & 6.5670   & 6.8279   \\
        119.088 & 100.3158 & 103.7812 & 107.7865
        \end{array}
\end{equation*}
All nominal median expected limits satisfy $\mu_{95}>1$. Consequently, none of the six benchmark signal normalizations would be excluded at the median expected 95\% confidence level under the assumptions of this analysis.  We repeat the likelihood fit using only the direct production signal while retaining the same classifier, score-bin boundaries, luminosity, background model, and flavor assumptions. The direct-only and combined results are summarized in Table~\ref{tab:combined_results}. 
\begin{table*}[htbp]
\centering
\begin{tabular}{c|ccc|c|cc|c}
\hline\hline
$m_N$ [GeV] & $N_D$ & $N_H$ & $N_{D+H}$ & $F_H$ [\%]
& $\mu_{95}^{D}$ & $\mu_{95}^{D+H}$ & $R_{95}$ \\
\hline
103.089 & 1.602 & 4.137 & 5.739 & 72.1 & 8.726 & 8.6271 & 1.0115 \\
103.355 & 9.782 & 24.382 & 34.164 & 71.4 & 1.562 & 1.5436 & 1.0119 \\
109.727 & 21.620 & 16.192 & 37.812 & 42.8 & 2.090 & 2.0681 & 1.0106 \\
112.174 & 4.548 & 3.697 & 8.245 & 44.8 & 3.309 & 3.2908 & 1.0055 \\
114.322 & 8.682 & 3.153 & 11.835 & 26.6 & 6.592 & 6.5670 & 1.0038 \\
119.088 & 0.928 & 0.162 & 1.090 & 14.9 & 103.968 & 103.7812 & 1.0018\\
\hline\hline
\end{tabular}
\caption{Direct and Higgs-mediated signal yields at
$1~\mathrm{ab}^{-1}$, together with the median expected
signal-strength limits from the direct-only and combined fits.}
\label{tab:combined_results}
\end{table*}
Here $N_D$, $N_H$, and $N_{D+H}$ are the Direct, Higgs-mediated, and combined signal yields, respectively. $F_H$ is the Higgs-mediated fraction of the selected signal, and $R_{95}=\mu_{95}^{D}/\mu_{95}^{D+H}$. Including the Higgs-mediated component improves the corresponding mixing limits by approximately $0.18$--$1.19\%$. This reflects the fact that the Higgs-mediated events populate classifier-score regions with comparatively lower sensitivity. We also validate the asymptotic approximation using 2000 pseudo-experiments for the $m_N=109.727~\mathrm{GeV}$ benchmark, evaluated at its median expected limit, $\mu_{95}=2.0681$. The asymptotic and toy-based calculations give $\mathrm{CL}_s=0.0501$ and $\mathrm{CL}_s=0.0462$, respectively. We increased the dominant $W\mu\nu$ background sample from $2\times10^{5}$ to $1.2\times10^{6}$ generated events to examine the sparsely populated classifier tail. No $W\mu\nu$ event entered the highest-score bin of the independent test sample. We therefore performed a conservative robustness test in which this bin was assigned an additional $W\mu\nu$ contribution of 27.1 events at $1~\mathrm{ab}^{-1}$. This corresponds to the one-sided $95\%$ Poisson upper bound for zero simulated survivors. For the nominal $10\%$ background uncertainty, the resulting median expected limits on the signal strength are $23.02$, $4.11$, $4.71$, $8.56$, $15.14$, and $\geq 200$, respectively, in order of increasing benchmark mass. We treat these as a robustness test of the high-score $W\mu\nu$ tail, not as our nominal result. Furthermore, we convert the expected limits on the signal strength into upper limits on the visible signal cross-section. Fig.~\ref{fig:visible_limits} shows the resulting limits for the six HNL mass hypotheses.
\begin{figure}[htbp]
    \centering
    \includegraphics[width=\columnwidth]
    {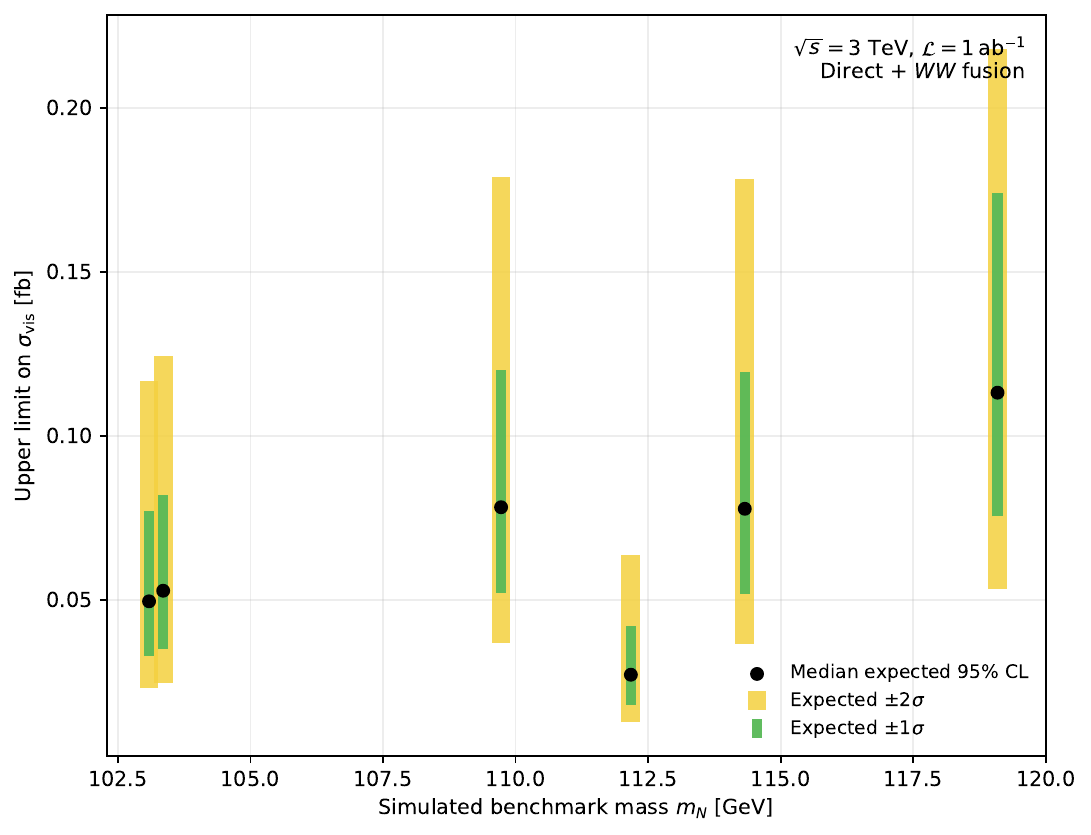}
    \caption{Median expected $95\%$ confidence-level upper limits
    on the visible signal cross section at the six HNL benchmark masses.}
    \label{fig:visible_limits}
\end{figure}
The mixing components are rescaled according to 
\begin{equation}
|U_{\alpha N}|^2=\kappa |U_{\alpha N}|^2_{\mathrm{BP}},
\end{equation}
We keep the flavor mixing ratios, branching fractions, and selection efficiencies fixed while the HNL mass is held fixed. In the limit $\Gamma(h\to N\nu)\ll\Gamma_h^{\mathrm{SM}}$, this gives $\kappa_{95}\simeq\mu_{95}$. Therefore, $|U_{\mu N}|^2_{95}\simeq\mu_{95}|U_{\mu N}|^2_{\mathrm{BP}}$.
The results are also interpreted as constraints on $|U_{\mu N}|^2$. Fig.~\ref{fig:physical_limits} shows the corresponding expected limits.
\begin{figure}[htbp]
    \includegraphics[width=\columnwidth]
    {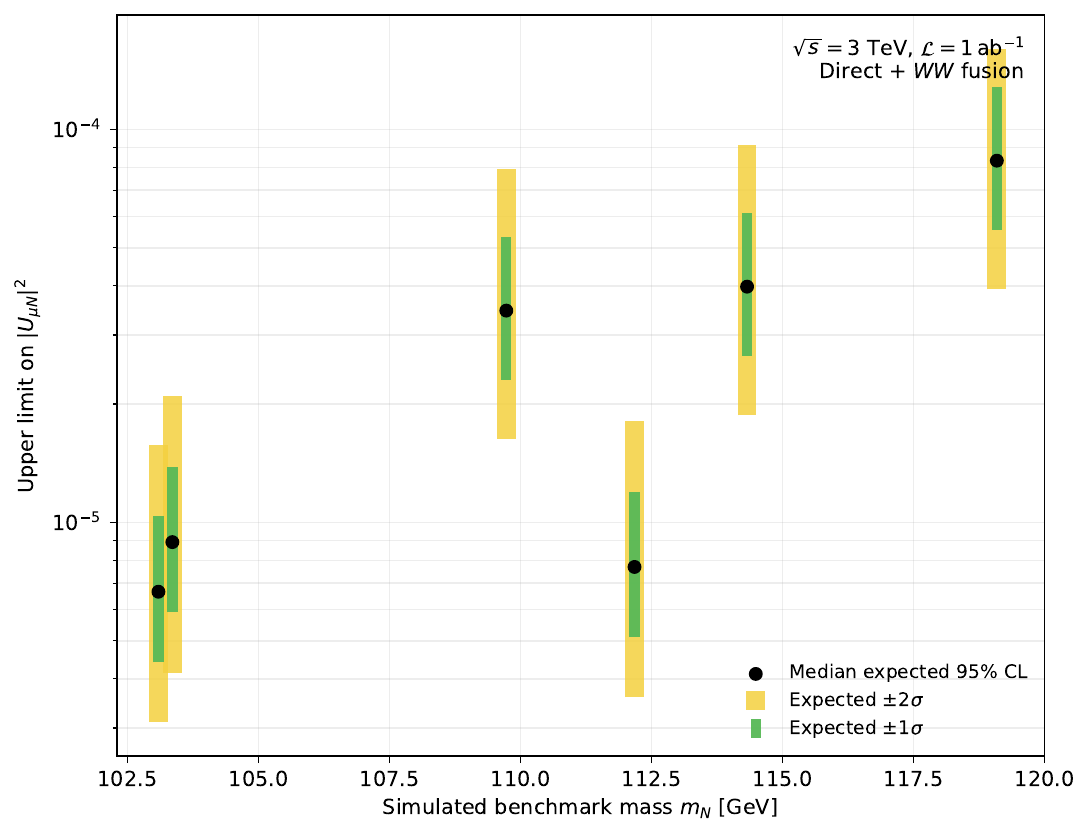}
    \caption{Median expected $95\%$ confidence-level upper limits on the active heavy neutrino mixing parameter $|U_{\mu N}|^2$ at the six HNL benchmark masses within the \texttt{T4-3-i-C1}.}
    \label{fig:physical_limits}
\end{figure}
The variation among the six limits is not only mass-dependent. The mixing parameters and branching fractions also differ between benchmark points. The smallest median signal-strength limit is $\mu_{95}=1.5436$ at $m_N=103.355~\mathrm{GeV}$. After accounting for the benchmark-dependent signal rates, the strongest visible-cross-section limit is approximately
$0.0271~\mathrm{fb}$ at $m_N=112.174~\mathrm{GeV}$. Also, after accounting for the benchmark-specific mixing parameters, the strongest limit on $|U_{\mu N}|^2$ is approximately $6.66\times10^{-6}$ at $m_N=103.089~\mathrm{GeV}$. For the $103.355$ and $109.727~\mathrm{GeV}$ benchmarks, the highest-score bin contains $\sim$ $3.18$ and $2.29$ expected signal events, respectively, over a nominal expected background of
$3.23$ events. At $119.088~\mathrm{GeV}$, the small $\mathrm{Br}(h\to N\nu)$, the selected yield of approximately one event, and only $0.046$ expected signal events in the highest-score bin lead to the weakest constraints. 
\section{Conclusion}
\label{sec:conclusion}
We investigated \textcolor{magenta}{the collider phenomenology of} a Dirac heavy neutral lepton (HNL) \textcolor{magenta}{with mass} below the Higgs-boson threshold in the \texttt{T4-3-i-C1} radiative neutrino mass model, 
\textcolor{magenta}{in which} direct and Higgs-mediated production are 
\textcolor{magenta}{controlled} by the same \textcolor{magenta}{underlying} model parameters. 
  \textcolor{magenta}{Focusing on} six benchmark points, we studied the common $\ell^\pm jj+p_{\mathrm{T}}^{\mathrm{miss}}$ signature at a $3~\mathrm{TeV}$ muon collider with \textcolor{magenta}{an integrated luminosity of}  $1~\mathrm{ab}^{-1}$. A detector-level analysis based on an \textsc{XGBoost} classifier and a binned profile likelihood fit \textcolor{magenta}{yields} median expected $95\%$ confidence-level limits of \textcolor{magenta}{approximately} $\sim0.027$--$0.113~\mathrm{fb}$ on the visible signal cross section; 
  \textcolor{magenta}{corresponding to limits of} $\times10^{-6}$--$8.3\times10^{-5}$ on $|U_{\mu N}|^2$ for the benchmark-specific flavor mixings. None of the nominal benchmarks would be excluded at the median expected $95\%$ confidence level.

The central result is that selected yield and exclusion sensitivity need not track one another. Although the Higgs-mediated channel contributes $\sim 15$--$72\%$ of the selected signal, its inclusion improves the expected mixing limits by only $0.18$--$1.19\%$ with the present classifier and binning. A dedicated Higgs-sensitive category may therefore extract this contribution more effectively. Further studies should also include additional background processes and beam-induced activity, improve the simulation of the high-score background tail, and extend the analysis beyond the six discrete benchmark points.

\section*{Acknowledgements}
The work of M.A.L. and S.N. is supported by the United Arab Emirates University (UAEU) under UPAR Grant No. 12S162.

\bibliographystyle{apsrev4-2} 
\bibliography{example}






\end{document}